\documentclass[twocolumn,showpacs,prl]{revtex4}
\usepackage{amsmath}
\usepackage{graphicx}
\usepackage{dcolumn}
\usepackage{bm}

\begin{document}

\title{ Tunneling density of states of granular metals}
\author{I.~S.~Beloborodov$^{1}$, A.~V.~Lopatin$^{1}$, G.~Schwiete$^{2}$,
and V.~M.~Vinokur$^{1}$}
\address{$^{1}$Materials Science Division, Argonne National
Laboratory, Argonne, Illinois 60439 \\ $^{2}$ Theoretische Physik
III, Ruhr-Universit\"{a}t Bochum, 44780 Bochum, Germany}

\date{\today}

\begin{abstract}
We investigate the effect of Coulomb interactions on the tunneling
density of states (DOS) of granular metallic systems at the onset
of Coulomb blockade regime in two and three dimensions. Using the
renormalization group technique we derive the analytical
expressions for the  DOS as a function of temperature $T$ and
energy $\epsilon.$ We show that samples with the bare
intergranular tunneling conductance $g^0_{\scriptscriptstyle T}$
less than the critical value $g_{\scriptscriptstyle
T}^{\scriptscriptstyle C}=(1/2\pi d) \ln(E_{\scriptscriptstyle
C}/\delta)$, where $E_{\scriptscriptstyle C}$ and $\delta$ are the
charging energy and the mean energy level spacing in a single
grain respectively, are insulators with a {\it hard gap} in the
DOS at temperatures $T\to 0.$ In 3d systems the critical
conductance $g_{\scriptscriptstyle T}^{\scriptscriptstyle C}$
separates insulating and metallic phases at zero temperature,
whereas in the granular films $g_{\scriptscriptstyle
T}^{\scriptscriptstyle C}$ separates insulating states with the
hard (at $g^0_{\scriptscriptstyle T}<g_{\scriptscriptstyle
T}^{\scriptscriptstyle C}$) and soft (at $g^0_{\scriptscriptstyle
T}>g_{\scriptscriptstyle T}^{\scriptscriptstyle C}$ ) gaps. The
gap in the DOS begins to develop at temperatures $ T^* \sim
E_{\scriptscriptstyle C} g_{\scriptscriptstyle
T}^{\scriptscriptstyle 0} \exp (-2\pi d g_{\scriptscriptstyle
T}^{\scriptscriptstyle 0})$ and reaches the value $\Delta \sim
T^*$ at $T\to 0$.

\end{abstract}

\maketitle

Granular materials, a focus of the current mesoscopic physics, is
a unique testing area for general concepts of disordered
systems~(see~\cite{experiment, Simon,Beloborodov99,Efetov02}). One
of the remarkable fundamental features of granular metals is the
strong on-site Coulomb repulsion that leads to the suppression of
transport at low temperatures and to an insulating ground state at
low bare intergranular conductances $g^0_{\scriptscriptstyle
T}<g_{\scriptscriptstyle T}^{\scriptscriptstyle C}=(1/2\pi d)
\ln(E_{\scriptscriptstyle C}/\delta)$. A good progress in
understanding this insulating state has recently been
made~\cite{Efetov02}; yet, despite the recent advance, the
satisfactory picture of physics of and near the Mott transition
that occurs at $g^0_{\scriptscriptstyle T}=g_{\scriptscriptstyle
T}^{\scriptscriptstyle C}$, was still lacking.  One of the
fundamental questions that remained open was the suppression of
the tunneling density of sates that always accompanies Mott
transition. In this Letter we develop a quantitative approach that
enables us to investigate Mott transition in granular metals and
derive the associated tunneling density of states in its vicinity.

The density of states is a fundamental quantity that determines
most of the properties of the system involved, and the electronic
transport is a key phenomenon where the manifestations of its
features may be most pronounced.  A general technique to treat
transport properties of granular metals in the high temperature
regime $T > g_{\scriptscriptstyle T} \delta$ was developed
recently in Ref.~\onlinecite{Efetov02}. It was, in particular,
shown that the conductivity  of granular metals can be described
in terms of the renormalized temperature dependent intergranular
tunneling conductance given by the following expression
    \begin{equation}
            \label{g} g_{\scriptscriptstyle T}(T)=
g_{\scriptscriptstyle T}^{\scriptscriptstyle 0}
            -(1/2\pi d)\,
\ln [g_{\scriptscriptstyle T}^{\scriptscriptstyle 0}\, E_{\scriptscriptstyle C}/T],
    \end{equation}
which holds as long as $g_{\scriptscriptstyle T}(T)> 1$. The
conductivity of the sample is related to the tunneling conductance
as $\sigma(T)=2 e^2 g_{\scriptscriptstyle T}(T) a^{2-d}$, where
$a$ is the granule size, $d$ is the dimensionality of the granular
array and the factor of 2 is due to the spin. From Eq.~(\ref{g})
follows that at temperatures
\begin{equation}
\label{T*} T^* \sim E_{\scriptscriptstyle C} g_{\scriptscriptstyle
T}^{\scriptscriptstyle 0}\, e^{-2 \pi d \, g_T^0 },
\end{equation}
the renormalized conductance, $g_{\scriptscriptstyle T}(T)$, is
strongly suppressed and approaches small values  where
renormalization group breaks down. Equation~(\ref{g}) is valid
only at temperatures $T>g_{\scriptscriptstyle T}
\delta$~\cite{Efetov} whereas in the opposite case,
$T<g_{\scriptscriptstyle T} \delta$, the conductance
renormalization~(\ref{g}) is saturated and the system behaves
essentially as a homogeneous disordered metal~\cite{BLV03}.
Comparing two relevant energy scales $T^*$ and
$g_{\scriptscriptstyle T} \delta$, one concludes~\cite{BLV03} that
if (i) $T^*<g_{\scriptscriptstyle T} \delta$ (or, equivalently,
$g_{\scriptscriptstyle T}^{\scriptscriptstyle 0}
> g_{\scriptscriptstyle T}^{\scriptscriptstyle C}$) then, the renormalized
conductance is still large at temperatures $T \sim
g_{\scriptscriptstyle T}\,\delta$ and the low temperature phase of
the system is similar to that of the disordered metals.
Alternatively, if (ii) $T^*> g_{\scriptscriptstyle T} \delta$ (or
$g_{\scriptscriptstyle T}^{\scriptscriptstyle
0}<g_{\scriptscriptstyle T}^{\scriptscriptstyle C}$), the
conductance of the system becomes significantly suppressed at
$T\sim T^*$ reflecting thus the onset of the Coulomb blockade
regime. In the latter case one expects that at $T\sim T^*$ the
Coulomb gap begins to develop (reaching its maximal value at zero
temperature) and, as a result, a noticeable suppression of DOS
even at finite temperatures, $T \sim T^*$ occurs.

In this Letter we consider the tunneling DOS of granular metals
with the bare tunneling conductance $g_{\scriptscriptstyle
T}^{\scriptscriptstyle 0} < g_{\scriptscriptstyle
T}^{\scriptscriptstyle C}$ at the onset of Coulomb blockade regime
at temperatures $T > T^{*}$ [case (ii) above]. We show that
Coulomb blockade strongly suppresses the tunneling DOS, $\nu(T)$.
For $3d$ granular samples $\nu(T)$ is given by
\begin{subequations}
\label{main}
\begin{equation}
\label{3d} \frac{\nu(T)}{\nu_{\scriptscriptstyle 0}}  = \left[ 1-
\frac{1}{6\pi g_{\scriptscriptstyle T}^{\scriptscriptstyle 0}}\ln
\frac{g_{\scriptscriptstyle T}^{\scriptscriptstyle
0}E_{\scriptscriptstyle C}}{T} \right]^{3A},
\end{equation}
whereas for  granular films we obtain
\begin{equation}
\label{2d} \frac{\nu(T)}{\nu_{\scriptscriptstyle 0}}  = \left[
{{g_{\scriptscriptstyle T}^{\scriptscriptstyle 0}
E_{\scriptscriptstyle C}}\over{ T}}\right]^{1/\pi} \left[1 -
\frac{1}{4\pi g_{\scriptscriptstyle T}^{\scriptscriptstyle 0}}\ln
\frac{g_{\scriptscriptstyle T}^{\scriptscriptstyle
0}E_{\scriptscriptstyle C}}{T}\right]^{4 g_{\scriptscriptstyle
T}^{\scriptscriptstyle 0}}.
\end{equation}
\end{subequations}
Here $\nu_{\scriptscriptstyle 0}$ is the DOS for non interacting
electrons and $A = 0.253$ is the dimensionless constant.
Equations~(\ref{main}) hold for temperatures $T > {\rm max} (T^*,
g_{\scriptscriptstyle T}\delta)$ where the temperature $T^*$ is
given by Eq.~(\ref{T*}). We note that according to Eqs.
(\ref{main}) the DOS vanishes exactly at the same temperature
$T^*$ as the renormalized conductance $g_T(T)$ in Eq.~(\ref{g}).

We show that the results~(\ref{main}) can be generalized to finite
frequency by substitution $T\to \rm{max} \{T,\epsilon\}$. In this
case Eqs.~(\ref{main}) can be applied even for $T\to 0$ provided
$\epsilon$ is larger than the characteristic energy $\Delta$ that
coincides with the temperature $T^*$
\begin{equation}
\label{Delta}
\Delta \sim E_{\scriptscriptstyle C} g_{\scriptscriptstyle
T}^{\scriptscriptstyle 0}\, e^{-2 \pi d \, g_{\scriptscriptstyle
T}^{\scriptscriptstyle 0} }.
\end{equation}

From Eqs.~(\ref{main}) one can see that the tunneling DOS is
strongly suppressed at energies $\epsilon \sim \Delta$. Thus, one
concludes that for $g_T^0<g_T^C$ the system is an insulator at
zero temperature with the Coulomb gap  $\Delta$ given by
Eq.~(\ref{Delta}),  see Fig.~1~a).
\begin{figure}[htb]
\includegraphics[width=1.0 \linewidth]{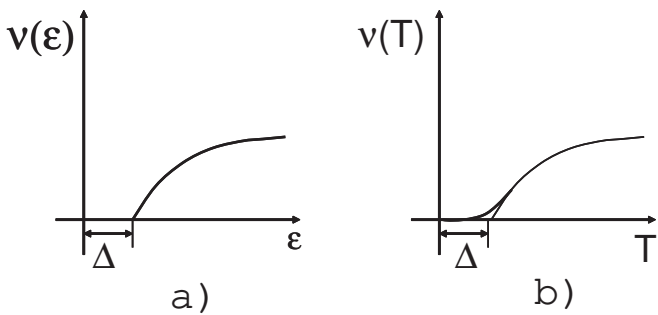}
\caption{Schematic behavior of the Density of States as a function
of a) energy and b) temperature.} \label{fig:2}
\end{figure}
On the contrary, for $\epsilon = 0$ but at finite temperatures,
$T$ we expect the exponentially small but finite activation
contribution for DOS
\begin{equation}
\nu(T)\sim \nu_0\, e^{-\Delta/T},
\end{equation}
see Fig.~1~b).

Equation~(1) generalizes straightforwardly to finite frequencies
by the substitution $T\to \rm{max} \{T,\omega\}$; this allows us
to relate the frequency dependent conductivity, $\sigma(\omega)$
with the tunneling density of sates at zero temperatures, $T=0$:
For $3d$ granular samples we obtain the following scaling relation
\begin{subequations}
\label{scaling}
 \begin{equation}
 \label{scaling3d}
 {{\nu(\omega)}\over {\nu_{\scriptscriptstyle 0}}}=\left[{{\sigma(\omega) }
 \over{\sigma_{\scriptscriptstyle 0}}}
 \right]^{3A},
 \end{equation}
 where $\sigma_{\scriptscriptstyle 0} = 2e^2g_T^0 a^{2-d}$
 is the high temperature conductivity. For
 granular films we get the following expression
 \begin{equation}
 \label{scaling2d}
{{\nu(\omega)}\over {\nu_{\scriptscriptstyle 0}}} = \left(
{{\Delta}\over{\omega }} \right)^{1/\pi}\, \left( {{
e\sigma(\omega) }\over{ \sigma_{\scriptscriptstyle 0}  }}
\right)^{4 g_{\scriptscriptstyle T}^{\scriptscriptstyle 0}},
 \end{equation}
 \end{subequations}
where $e \approx 2.7182.$
Equations~(\ref{scaling}) are useful for the comparison of our
predictions with the experimental data.

Now we turn to the quantitative description of our model and
derivation of Eq.~(\ref{main}):  We consider a $d-$dimensional
array of metallic grains. The motion of electrons inside the
grains is diffusive and they can tunnel between grains. We assume
that in the absence of the Coulomb interaction, the sample would
be a good metal.

The system of weakly coupled metallic grains is described by the
Hamiltonian
\begin{subequations}
\label{hamiltonian}
\begin{equation}
\hat{H} = \hat{H}_{0} + \hat{H}_{c} + \hat{H}_{t}.
\label{hamiltonian1}
\end{equation}
The term $ \hat{H}_{0}$ in Eq.~(\ref{hamiltonian1}) is the
Hamiltonian of an isolated disordered granule. The term
$\hat{H}_{c}$ describes the Coulomb repulsion  and is given by
\begin{equation}
\label{Hc} \hat{H}_{c}={\frac{{\ e^{2}}}{{\
2}}}\,\sum_{ij}\,\hat{n}_{i}\,C_{ij}^{-1}\, \hat{n}_{j},
\end{equation}
where $i$ stands for the granule number, $C_{ij}$ is the
capacitance matrix and $\hat{n}_{i}$ is the operator of electrons
number in the $i$-th granule. The last term on the right hand side
of Eq.~(\ref{hamiltonian1}) is the tunneling Hamiltonian
\begin{equation}
\label{Ht} \hat{H}_{t} = \sum_{ij,p,q} t_{ij} a^{\dagger}_{i,p}
a_{j,q},
\end{equation}
\end{subequations}
where $a^{\dagger}_{i,k} (a_{i,k})$ are the creation
(annihilation) operators for an electron in the state $k$ of the
$i$-th grain and $t_{ij}$ is the tunneling matrix element
corresponding to the points of contact of $i$-th and $j-$th
granules.

As it was shown in Ref.~\onlinecite{Efetov}, at temperatures $T >
g_{\scriptscriptstyle T}\delta$ the model given by
Eq.~(\ref{hamiltonian}) can be effectively described in terms of
the functional proposed by Ambegaokar, Eckern and Sch\"{o}n (AES)
in Ref.~\cite{AES}:
\begin{subequations}
\begin{equation}
\label{action} S = S_c + S_t,
\end{equation}
where $S_c$ is the charging energy
\begin{equation}
S_c = \sum\limits_{ij} \int_{0}^{\beta}d\tau
\frac{d\phi_i}{d\tau}\frac{C_{ij}}{2e^2}\frac{d\phi_j}{d\tau},
\end{equation}
and the second term in the right hand side of Eq.~(\ref{action})
describes tunneling between the granules
\begin{equation}
S_t = 2\pi g_T \sum\limits_{ij}\int_0^{\beta}\frac{T^2 d\tau
d\tau'}{ \sin^2[\pi T(\tau -
\tau')]}\sin^2\left[\frac{\phi_{ij}(\tau) -
\phi_{ij}(\tau')}{2}\right].
\end{equation}
\end{subequations}
Here $\phi_{ij}(\tau) = \phi_i(\tau) - \phi_j(\tau)$ is the phase
difference between the phase in the i-th and j-th granules. In the
metallic regime we may neglect winding numbers in the phases. We
use the renormalization group technique to calculate DOS.  The
charging part, $S_c$ of the action in Eq.~(\ref{action})
determines the upper frequency cutoff.

In terms of the functional approach based on the action
(\ref{action}), DOS is expressed as~\cite{Efetov02}
\begin{equation}  \label{nu}
\frac{\nu(\epsilon)}{\nu_0}= T \, {\rm  Im} \,  \int_0^\beta
{{d\tau e^{i\epsilon_n\tau }\over {\sin \pi T\tau }} \langle
e^{-i(\phi_i(\tau)-\phi_i(0)} ) } \rangle
_{\epsilon_n\to-i\epsilon }.
\end{equation}
Following the standard renormalization group procedure we separate
the field $\phi$ into slow, $\phi_s$ and fast, $\phi_f$ parts and
integrate over the fast field in Eq.~(\ref{nu}). The fast fields
belong to the interval $dS$ that represents the energy shell
$d\Lambda$ in the configuration space of the quasimomentum, ${\bf
q}$ and frequency, $\omega$. Using the one-loop approximation we
obtain the following RG equation for the conductance $g_T$
\cite{Efetov02}
\begin{equation}
{{ d g_T(\Lambda)}\over {d \Lambda}} = {{1}\over {2\pi d \Lambda
}}. \label{rg_conductance}
\end{equation}
Performing the integration in Eq.~(\ref{rg_conductance}) we obtain
Eq.~(\ref{g}). With the same accuracy for the flow equation of the
density of sates we get
\begin{equation}
\label{RG} d \,\ln(\nu/\nu_0) =  a^d \int_{dS}
{{d\omega}\over{2\pi }}\, {{ d^d q }\over{(2\pi)^d }}
G_\phi(\omega,q).
\end{equation}
Here the Green function $G_\phi(\omega,q)$ of the phase fields
$\phi$ is defined on the scales $\Lambda$
\begin{equation}
\label{G}
 G_{\phi}(\omega,q)={1\over{2 g_T(\Lambda) }} \,
{{1}\over {|\omega| E_q}}.
\end{equation}
In equation~(\ref{G}) we introduced the notation $E_q=2
\,\sum_{\bf a}[1-\cos({\bf q a})].$ The integration in
Eq.~(\ref{RG}) is going over the infinitesimal volume $dS$ in the
($\omega,{\bf q}$) configuration space that corresponds to the
energy interval $d\Lambda$. The proper way to chose a particular
form of $dS$ depends on the dimensionality of the sample: For $3d$
samples the integrals over the quasimomentum converge and one can
simply choose $dS= (2\pi/a)^3 d\Lambda$. This leads to the
following differential equation
\begin{equation}
\label{dif_eq1}
 d \ln(\nu/\nu_0)= {{ A }\over{ 2\pi }} {{ d
\Lambda}\over {\Lambda \, g_{\scriptscriptstyle T}(\Lambda) }},
\end{equation}
where
 $
 A=a^3 \int {{ d^3 q}\over { (2\pi)^3}}
 {{1}\over { E_q}} \approx 0.253
 $
is the numerical constant. Integrating over the $\Lambda$ in
Eq.~(\ref{dif_eq1}) in the range ($T, g_{\scriptscriptstyle
T}^{\scriptscriptstyle 0}\, E_{\scriptscriptstyle C}$) we obtain
Eq.~(\ref{3d}) for the DOS of $3d$ granular metals.

The $2d$ case is different since the direct integration over the
quasi-momentum, ${\bf q}$ in Eq.~(\ref{RG}) would lead to the
infra-red divergence. In this case it is natural to introduce the
infinitesimal volume, $dS$ in the following way
\begin{equation}
\label{ds} \int_{dS}\, d\omega d^2 q = \int \, d\omega\, d^2 q \,
\delta(|\omega| E_q -\Lambda)  d\Lambda,
\end{equation}
such that on the energy shell $\Lambda$ the propagator (\ref{G})
will not be divergent since $\omega E_q=\Lambda$. Using
Eq.~(\ref{ds}), performing the integration  over the $\omega$, $q$
in Eq.~(\ref{RG}) and taking into account the fact that the upper
cutoff for frequency, $\omega$ is $g_{\scriptscriptstyle T}\,
E_{\scriptscriptstyle C}(q),$ where for $2d$ granular samples the
charging energy, $E_{\scriptscriptstyle C}(q)$ is given by the
expression $E_{\scriptscriptstyle C}(q)=\pi E_c /q a$, with the
logarithmic accuracy we obtain the following equation
\begin{equation}
\label{RG_2} d \ln(\nu/\nu_0)=  {{1}\over{4\pi^2 }} {{ 1 }\over
{\Lambda \, g_{\scriptscriptstyle T}(\Lambda)  }} \,
\ln\left[\frac{g_{\scriptscriptstyle T} E_{\scriptscriptstyle
C}}{\Lambda}\right]\, d\Lambda .
\end{equation}
Integrating Eq.~(\ref{RG_2}) over the variable $\Lambda$ in the
interval $(T, g_{\scriptscriptstyle T} E_{\scriptscriptstyle C})$
we obtain DOS as given by Eq.~(\ref{2d}) for $2d$ granular
samples. Equation~(\ref{2d}) was derived for long range Coulomb
interaction. For short range Coulomb interaction (this may happen
due to the presence of external screening) the upper cutoff for
$\omega$-integration in Eq.~(\ref{RG}) is $q-$ independent and is
given by $g_{\scriptscriptstyle T} E_{\scriptscriptstyle C}$. In
this case the extra factor $1/2$ will appear  in the right hand
side of Eq.~(\ref{RG_2}) and as a consequence the final result for
the DOS in Eq.~(\ref{2d}) will be modified as $
\nu(\omega)/\nu_{\scriptscriptstyle 0} \to [
\nu(\omega)/\nu_{\scriptscriptstyle 0}]^{1/2}$. In this case for
$g_{\scriptscriptstyle T}^{\scriptscriptstyle 0} >
g_{\scriptscriptstyle T}^{\scriptscriptstyle C}$ one reproduces
the result of Ref.~\cite{Efetov02} for the tunneling DOS of
granular metals in the limit of large tunneling conductance.

Although our theory applies to $1d$ granular arrays, the results
in this case should be taken with caution when compared with
experimental data: the problem is that the conductivity of $1d$
system is usually controlled by the weakest junction and thus must
be described by the conductance distribution function~\cite{LV}
rather than by the average value of the conductance. Thus even
small fluctuations of conductance could be important in $1d$ case,
especially close to the predicted transition at
$g^0_{\scriptscriptstyle T}\sim g_{\scriptscriptstyle
T}^{\scriptscriptstyle C}$; we, however leave the detailed
analysis of this situation to the forthcoming publication.

In conclusion, we have investigated the effect of Coulomb blockade
on the tunneling DOS of granular metals in the limit of large
tunneling conductance between the grains. We have determined the
critical value of tunneling conductance $g_{\scriptscriptstyle
T}^{\scriptscriptstyle C}=(1/2\pi d) \ln (E_{\scriptscriptstyle
C}/\delta)$ below which the granular metal becomes an insulator
with a ``hard'' gap at zero temperature. For $3d$ samples this
value of critical conductance corresponds to  a metal-insulator
transition, as granular samples with $g_{\scriptscriptstyle
T}^{\scriptscriptstyle 0}>g_{\scriptscriptstyle
T}^{\scriptscriptstyle C}$ are metallic at zero
temperature~\cite{BLV03}. This value of $g_{\scriptscriptstyle
T}^{\scriptscriptstyle C}=(1/2\pi d) \ln (E_{\scriptscriptstyle
C}/\delta)$ explains the long known puzzling fact that in $3d$
systems the metal insulator transition occurs at
$g_{\scriptscriptstyle T}\approx 0.1$.

The situation is different for $2d$ granular systems since in this
case even samples with $g_{\scriptscriptstyle
T}^{\scriptscriptstyle 0}>g_{\scriptscriptstyle
T}^{\scriptscriptstyle C}$ are insulators at temperatures  $T\to
0$ due to interaction and quantum effects~\cite{BLV03}, similar to
those that take place in homogeneously disordered metals~\cite{AA,
Finkelstein,Shklovskii}. Nevertheless even in $2d$ case the
critical value of conductance $g_T^C$ represents  the boundary
between two physically different regimes at temperatures $T\to 0$:
Samples with $g_{\scriptscriptstyle T}^{\scriptscriptstyle
0}<g_{\scriptscriptstyle T}^{\scriptscriptstyle C}$ represent the
``hard'' insulators, with a hard gap in the DOS, while samples
with $g_{\scriptscriptstyle T}^{\scriptscriptstyle
0}>g_{\scriptscriptstyle T}^{\scriptscriptstyle C}$ are insulators
with a soft gap in the DOS similar to homogeneously disordered
metals.

We like to thank K.~Efetov and Yu.~Galperin for useful
discussions. This work was supported by the U.S. Department of
Energy, Office of Science through contract No. W-31-109-ENG-38.
G.~S. thanks Material Science Division of Argonne National Laboratory for
hospitality and greatfully acknowledges the support of GRK 384.

\end{document}